\documentclass[twoside,slac_one]{revtex4}
\usepackage{graphicx}
\usepackage{fancyhdr}
\usepackage{amsmath} % American Mathematics Society standards
\usepackage{bm}% bold math
\usepackage{amsxtra}
\usepackage{amssymb}
\usepackage{amsthm}
\usepackage{latexsym}
\usepackage{lscape}

\begin{document}

\title{Searches for Large Extra Dimensions at CMS}


\author{Alexey V. Ferapontov (for the CMS Collaboration)}
\affiliation{Department of Physics and Astronomy, Brown University, Providence, RI, USA}

\begin{abstract}
Results of searches for large extra dimensions in $pp$ collisions at the center-of-mass energy of 7~TeV with the CMS detector are presented. No excess of events above the standard model expectations in up to 1.1~fb$^{-1}$ of data is found. Stringent limits are set on the multidimensional Planck scale as well as masses of exotic objects that are consequences of the extra dimensions.
\end{abstract}

\maketitle

\thispagestyle{fancy}


%%%%%%%%%%%%%%%%%%%%%%%%%%%%%%%%%%
\section{Introduction\label{s:intro}}
In our (3+1)-dimensional Universe, gravity is the weakest of all forces~\cite{SM1,SM2}. The fact that the electroweak scale ($\mathcal{O}$(0.1 -- 1~TeV)) is so much lower than the scale when gravity becomes strong (known as the Planck scale and is $\mathcal{O}$(10$^{16}$~TeV)) is known as the hierarchy problem~\cite{hierarchy}. 

In this Letter, two different theories that attempt to solve or alleviate the hierarchy problem are discussed, with both theories involving extra dimensions. The first model, proposed by Arkani-Hamed, Dimopoulos, and Dvali (ADD model)~\cite{tevscale, bhtheory}, postulates that gauge interactions are localized on a (3+1) brane and gravity propagates in $n$ large, flat extra dimensions, compactified on a torus or sphere with radius $R$. The multidimensional Planck scale ($M_D$) is related to fundamental Planck scale ($M_{Pl}$) as: $M_{Pl}^{2} = 8 \pi M_D^{n+2} R^n$, where $R$ is the size of extra dimensions. For $M_D = 1$~TeV (or in other words, comparable to electroweak scale), $R$ can be as large as $10^{13}$~m for $n = 1$ and rapidly decreases to 3~nm for $n = 3$, $6\times 10^{-12}$~m for $n = 4$, and so on.
 
Another possible solution of the hierarchy problem is due to a virtual graviton exchange in extra dimensions~\cite{add_diphot,grw,hlz,hewett}. The result is a non-resonant production of dileptons or diphotons at high masses, that appear as a continuum spectrum of Kaluza-Klein (KK) excitations separated by energy gaps inversely proportional to the size of extra dimensions, $R$. Typical values of $R^{-1}$ for number of extra dimensions of 2, 4, and 6 are 40~meV, 40~keV, and 10~MeV, respectively. The summation over all KK modes diverges, hence, an ultraviolet cutoff (model-dependent) is required. There are more than one possible scenarios~\cite{add_diphot,grw,hlz,hewett}, in which the effects of the virtual graviton production are parameterized differently -- this will be described in more details in the respective Sections of this Letter. Finally, a resonant production of diphotons by means of Randall-Sundrum (RS) graviton decay is possible~\cite{rs}. In this model, gravity appears strengthened in only one extra dimension (RS-1 model) due to its warped geometry. The model parameters are the curvature of the metric ($k$), mass of the first excitation ($M_1$), and the ultraviolet cutoff scale ($M_S$). Often, a dimensionless parameter $\tilde{k} = k/M_{Pl}$ is used instead of the curvature.

The searches are carried out by the Compact Muon Solenoid (CMS) Collaboration using $pp$ collisions at the center-of-mass energy $\sqrt{s} = 7$~TeV at the Large Hadron Collider (LHC). The detailed description of the CMS detector can be found elsewhere~\cite{CMS}. The data for the searches described in this Letter have been taken during the 2010 and 2011 LHC runs and correspond to an integrated luminosity of up to 1.1~fb$^{-1}$~\cite{lumi}. 

%%%%%%%%%%%%%%%%%%%%%%%%%%%%%%%%%%
\section{Searches for Microscopic Black Holes\label{s:bh}}
One of the consequences of a strengthened gravitational field in extra dimensions is possibility of microscopic black holes formation if the impact parameter of two colliding particles is smaller then a $(4+n)$-dimensional Schwarzschild radius ($r_S$)~\cite{mp1,mp2,mp3}:
$r_S = \frac{1}{\sqrt \pi M_D} {\left[ \frac{M_{\rm BH}}{M_D} \frac{8\Gamma\!\left(\frac{n + 3}{2}\right)}{n + 2} \right]}^{\frac{1}{n + 1}},
$
where $M_{\rm BH}$ is the mass of a black hole. The production cross section of such black holes is $\sigma \sim \pi r_{S}^{2}$ and can reach hundreds of picobarns, which makes search for such objects promising at the LHC. In our current understanding of a model, in the semiclassical approximation, the minimum allowed mass of a black hole $M_{\rm BH}^{\rm min}$ is some 3 -- 5 times larger than the $M_{D}$, although the formation threshold can be even larger. In cases when the semiclassical approximation no longer holds, string balls~\cite{sb} may offer a better description of the formation of a black hole. String balls are considered to be precursors of black holes, with mass being close to the Planck scale. They are described by the string scale $M_S$ and the string coupling $g_S$. The properties of string balls are similar to that of black holes: once produced, they both decay thermally into all standard model (SM) degrees of freedom. Since quarks and gluons carry color charge, they are the main decay products ($\sim 75\%$). 

On average, half a dozen energetic particles is expected in the final state, which drives the analysis strategy. This search~\cite{exo-11-071} is based on $1.09 \pm 0.07$~fb$^{-1}$ of collision data, and is an update of two previous searches~\cite{bh_plb,exo-11-021}. Events are pre-selected based on the total jet activity triggers with scalar sum of central (with pseudorapidity $-3.0 < \eta < 3.0$, where $\eta = -\ln[\tan(\theta/2)]$, $\theta$ is the polar angle measured from the nominal detector center with respect to the direction of the counterclockwise proton beam) jets transverse energies ($E_T$) between 350 and 550~GeV. Offline, $S_T$ is the main variable that is used to discriminate the signal and the background. It is defined as a scalar sum of the $E_T$ (or transverse momentum $p_T$ for muons) of all isolated and non-overlapping jets, photons, electrons, and muons that pass loose object identification requirements and have $E_T > 50$~GeV. If missing transverse energy ($\ensuremath{E_{T}^{\rm{miss}}}$) is greater than 50~GeV, it is added to the $S_T$. The total number of reconstructed objects (excluding $\ensuremath{E_{T}^{\rm{miss}}}$) is called multiplicity, $N$, and is used in the offline selection. To avoid trigger and multi-particle turn-ons, events with $S_T < 800$~GeV are rejected. Note that by construction, particle identification efficiency does not affect the $S_T$ distribution, since if an electron does not pass the identification requirements it is either classified as a photon, or as a jet; photon failing the selection will become a jet; and a rejected muon will contribute to $\ensuremath{E_{T}^{\rm{miss}}}$.

\begin{figure}[htbp]
\includegraphics[scale=0.25]{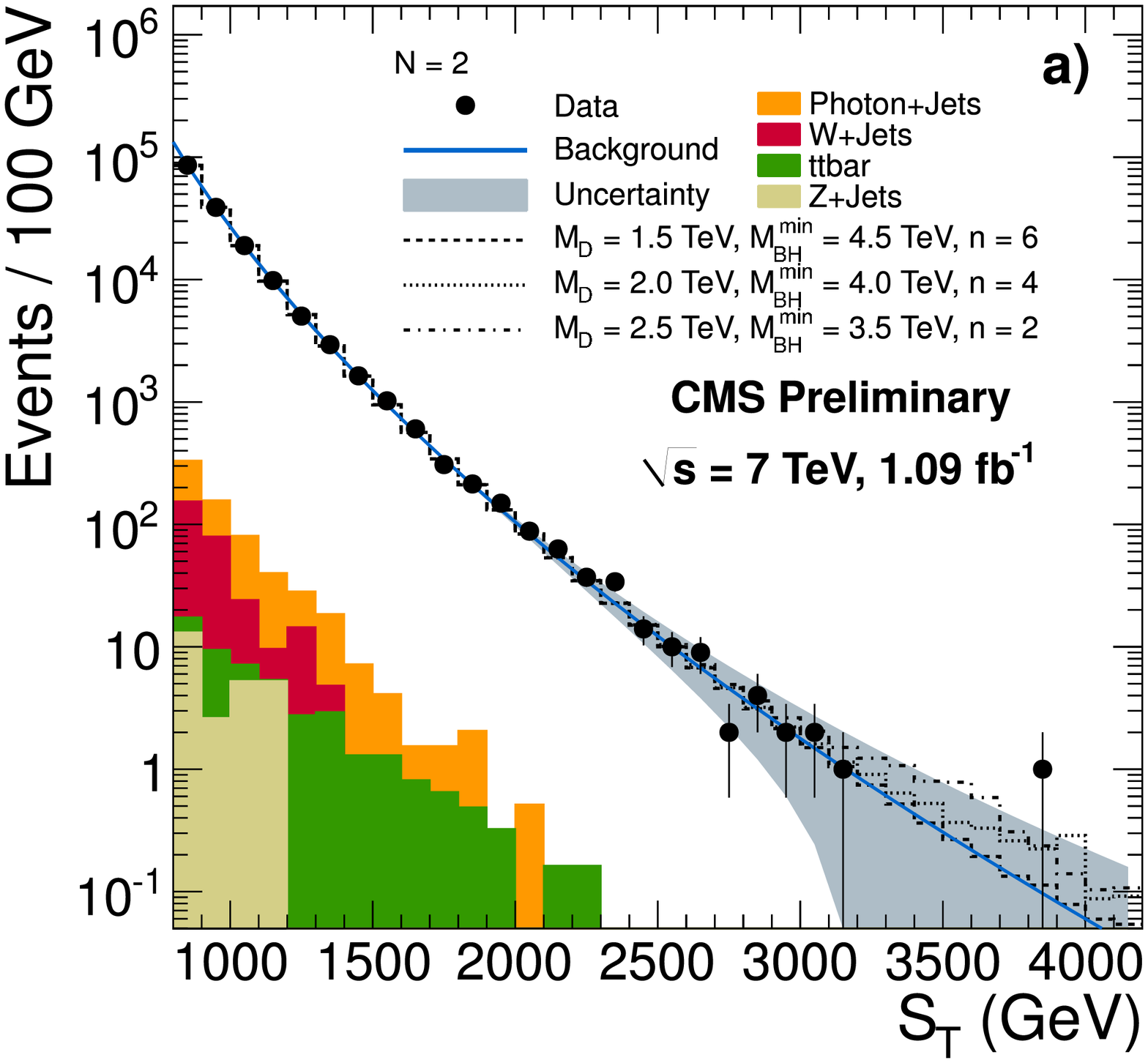}
\includegraphics[scale=0.25]{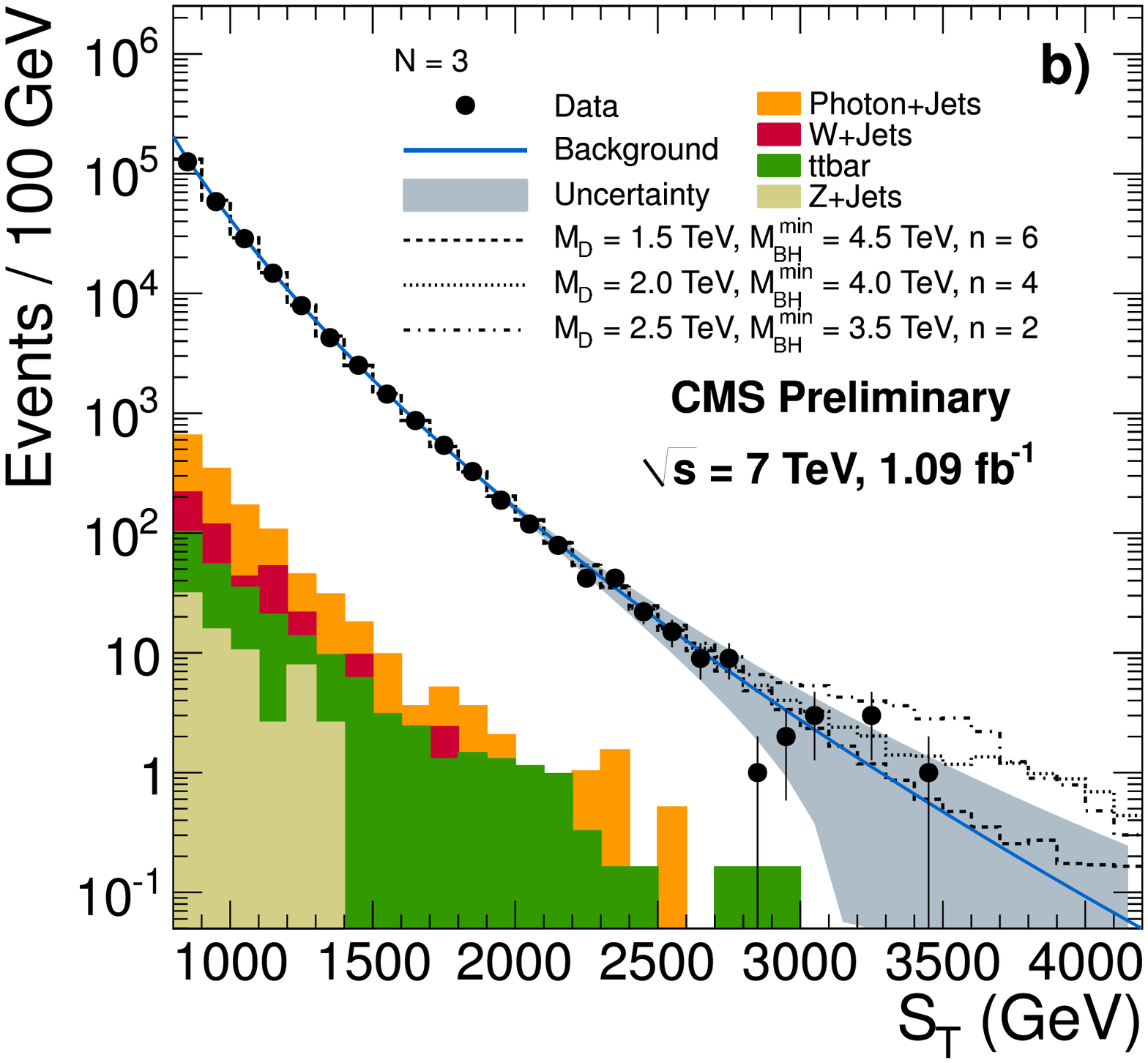}\\
\includegraphics[scale=0.25]{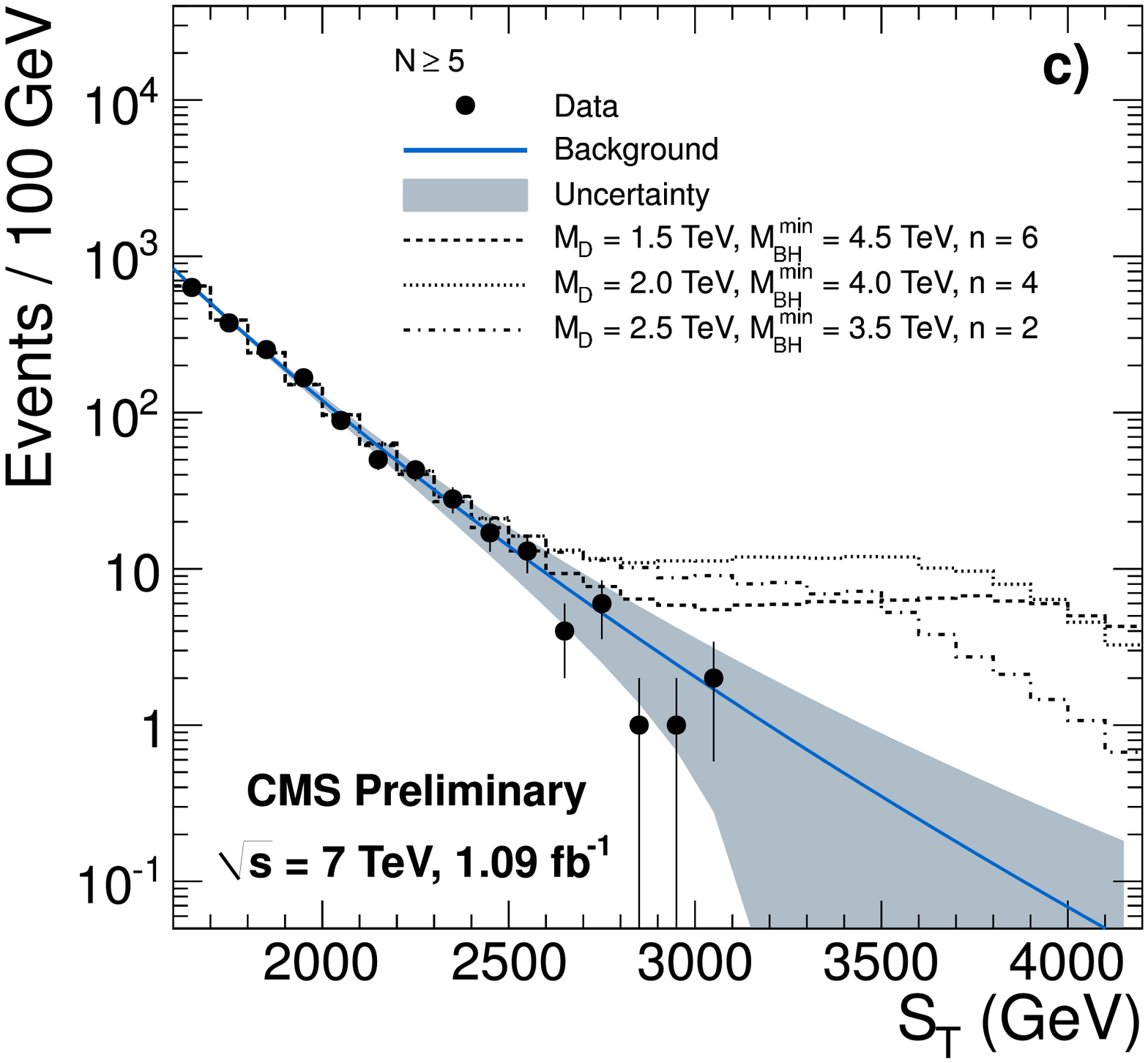}
\includegraphics[scale=0.25]{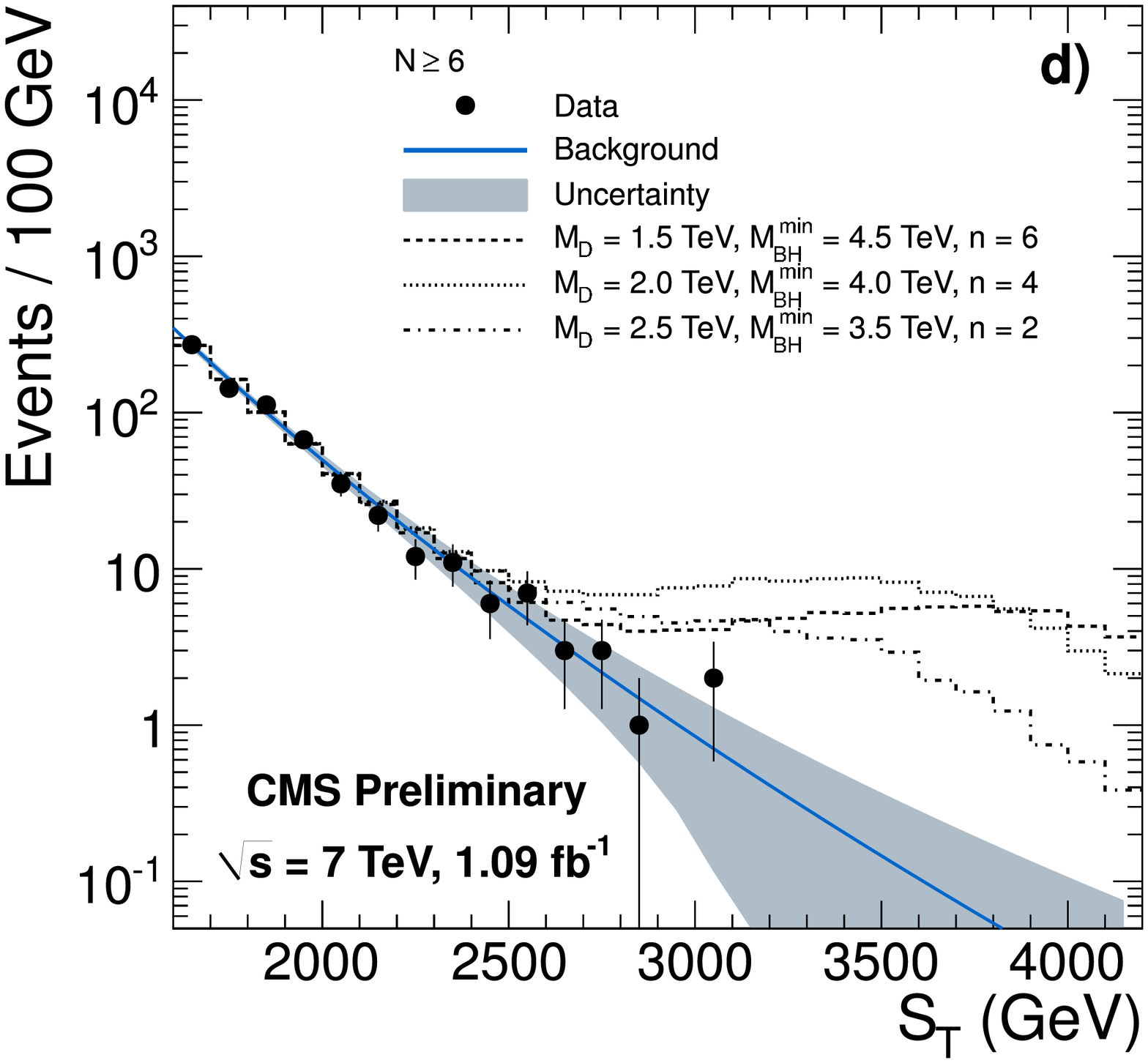} 
\caption{Total transverse energy, $S_T$, for events with a) $N = 2$, b) $N = 3$, c) $N \ge 5$, and d) $N \ge 6$ objects in the final state. Data are depicted as solid circles with error bars; shaded band is the data-driven background prediction (solid line) with its uncertainty. Non-QCD backgrounds are shown as colored histograms. Also shown is black hole signal for three different parameter sets.
\label{fig:st_excl}}
\end{figure}

The backgrounds to the black hole signals include $V$+jets (where $V = W, Z, \rm{or}~\gamma$) and multijet QCD production. The former backgrounds are negligibly small at high $S_T$, and are omitted in the analysis. The dominant QCD background is estimated using a data-driven technique that is based on the assumption that the $S_T$ shape is invariant of the multiplicity of the final state. The $S_T$ distributions in data for $N = 2$ (almost completely dominated by hard QCD $2 \rightarrow 2$ scattering) and $N = 3$ are fitted between 800 and 2500~GeV (where no signal is expected) with three ansatz functions: the one with the best $\chi^{2}$/n.d.f. becomes a template for the background, and the other two are used to derive the systematic uncertainties due to the method (see Fig.~\ref{fig:st_excl}). Further, the template function is normalized to inclusive multiplicities in data between 1.6 and 2.0~TeV, and extrapolated to the search region. No  significant excess of data over the background prediction in all inclusive multiplicities from $N \ge 3$ to $N \ge 8$ is found. For the complete set of figures, see Ref.~\cite{exo-11-071}, and in this Letter only $N \ge 5$ and $N \ge 6$ plots are shown (see Fig.~\ref{fig:st_excl}).

\begin{figure}[htbp]
\includegraphics[scale=0.25]{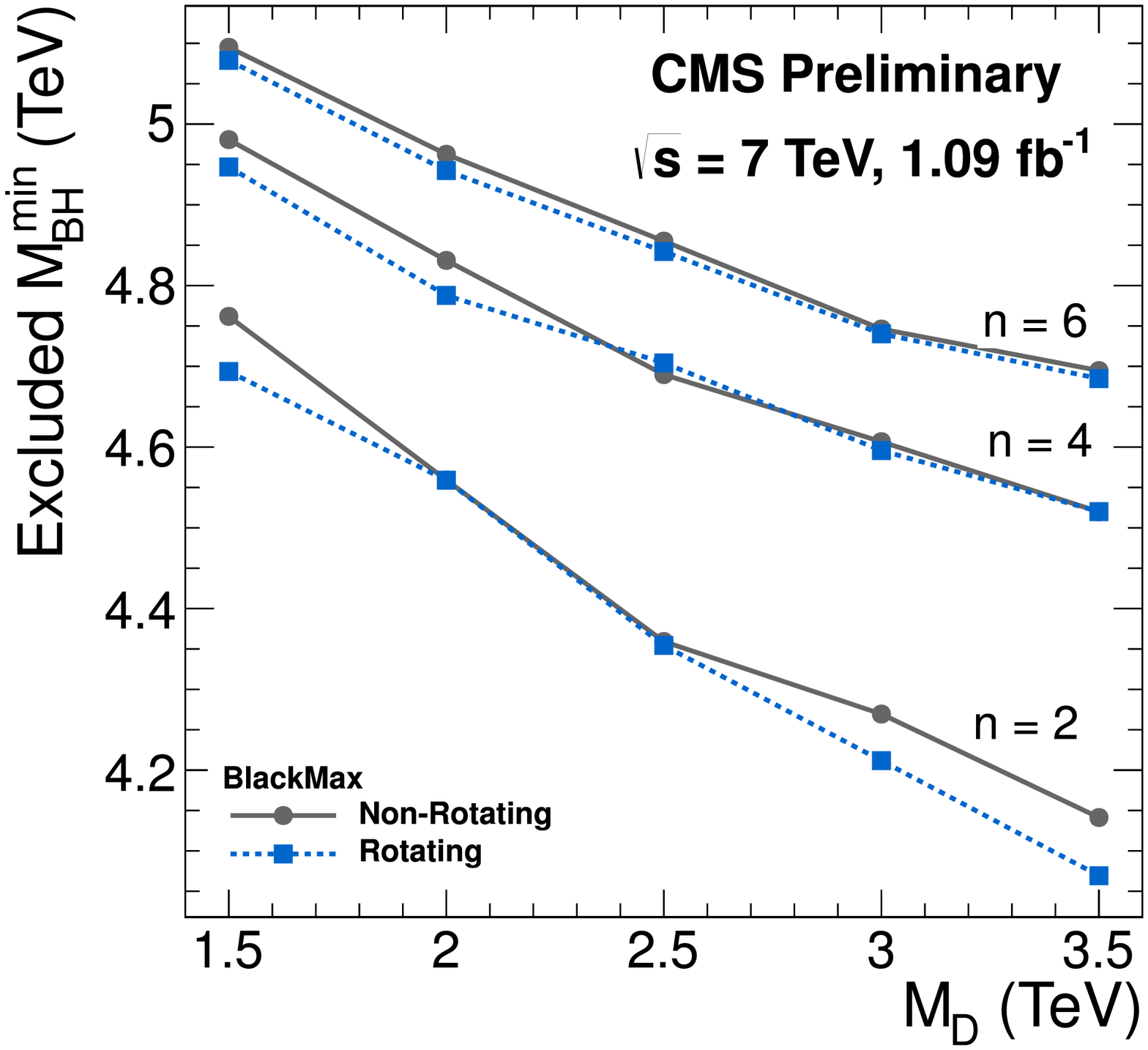}
\includegraphics[scale=0.25]{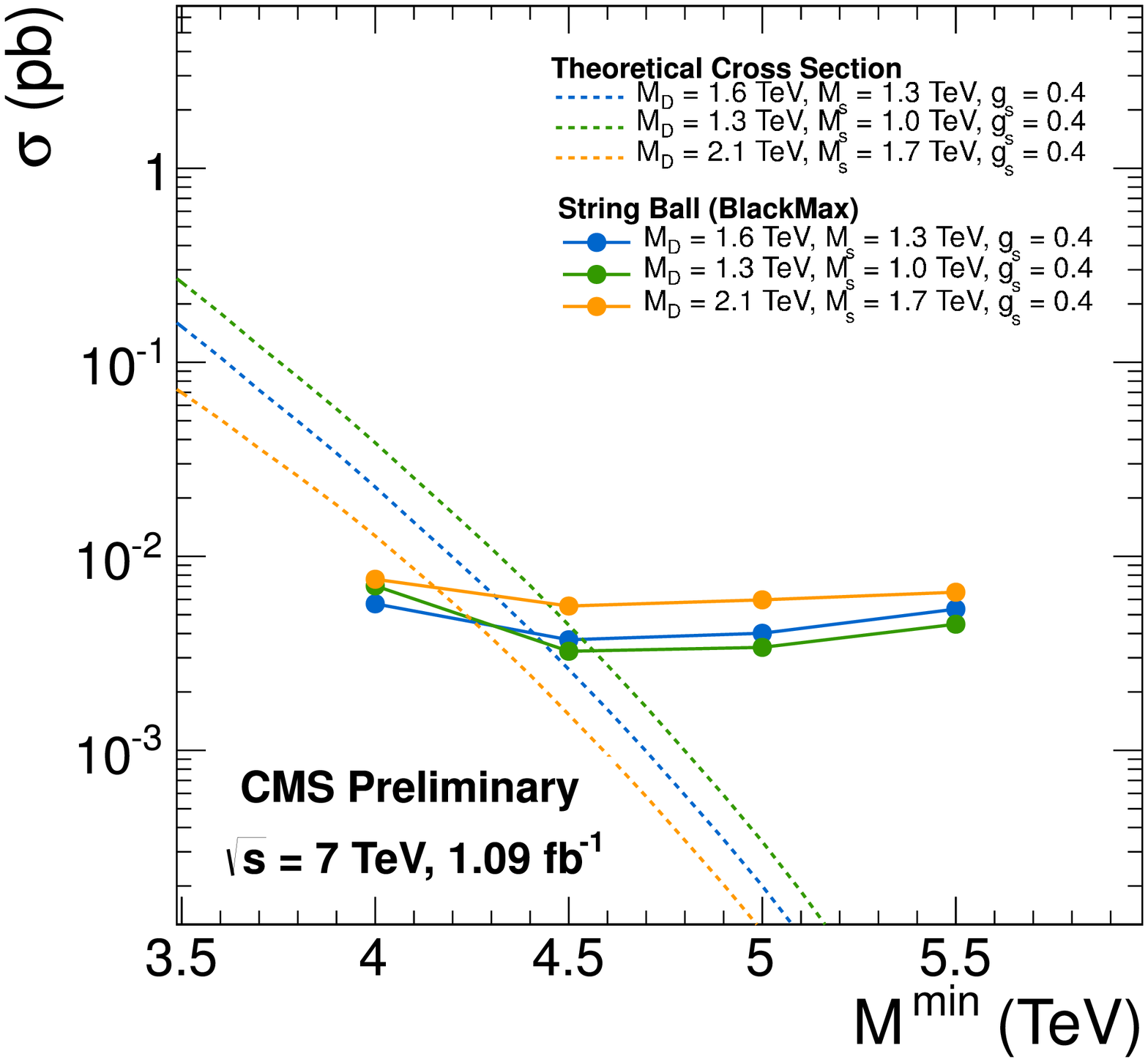}\\
\includegraphics[scale=0.25]{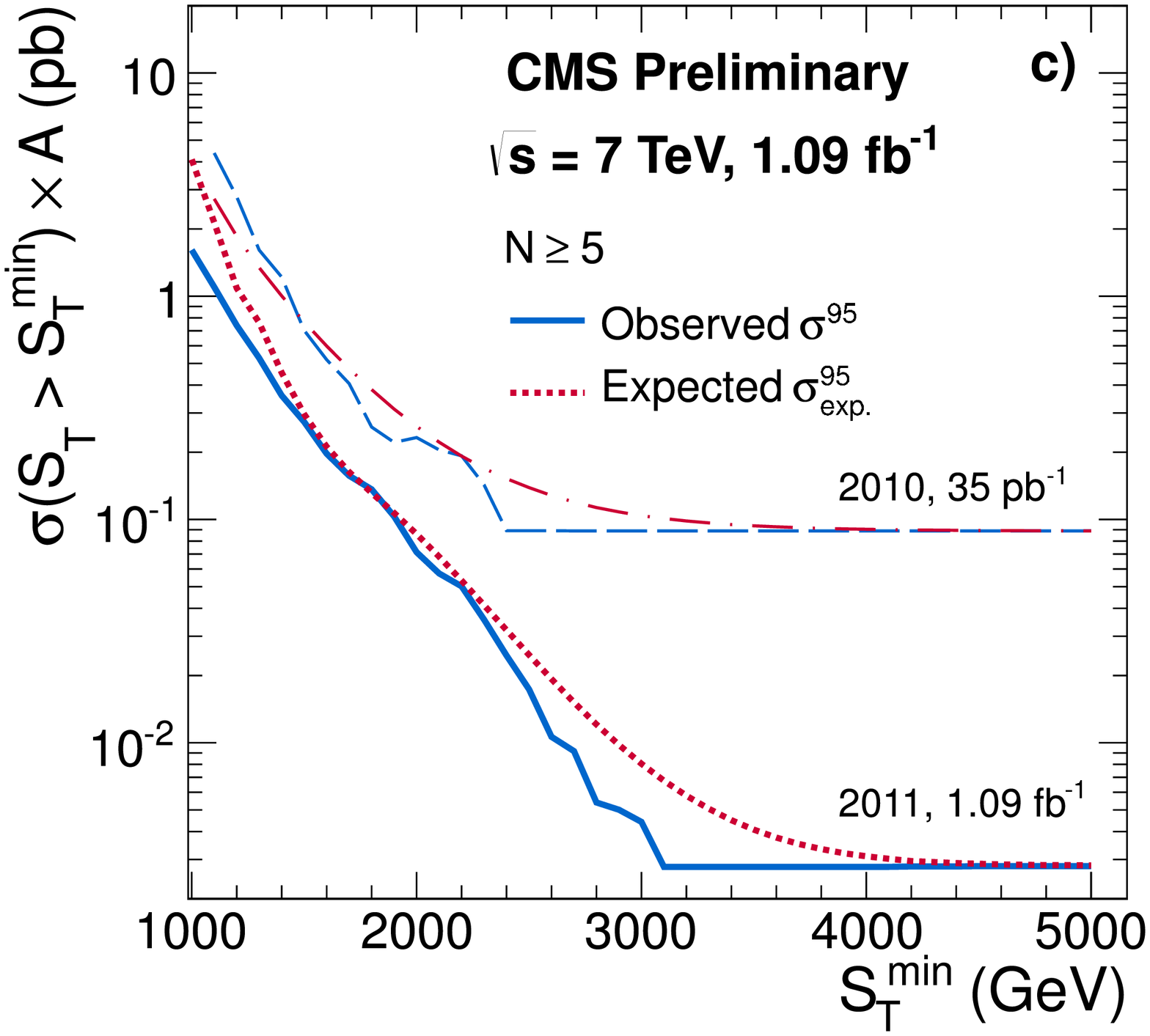}
\includegraphics[scale=0.25]{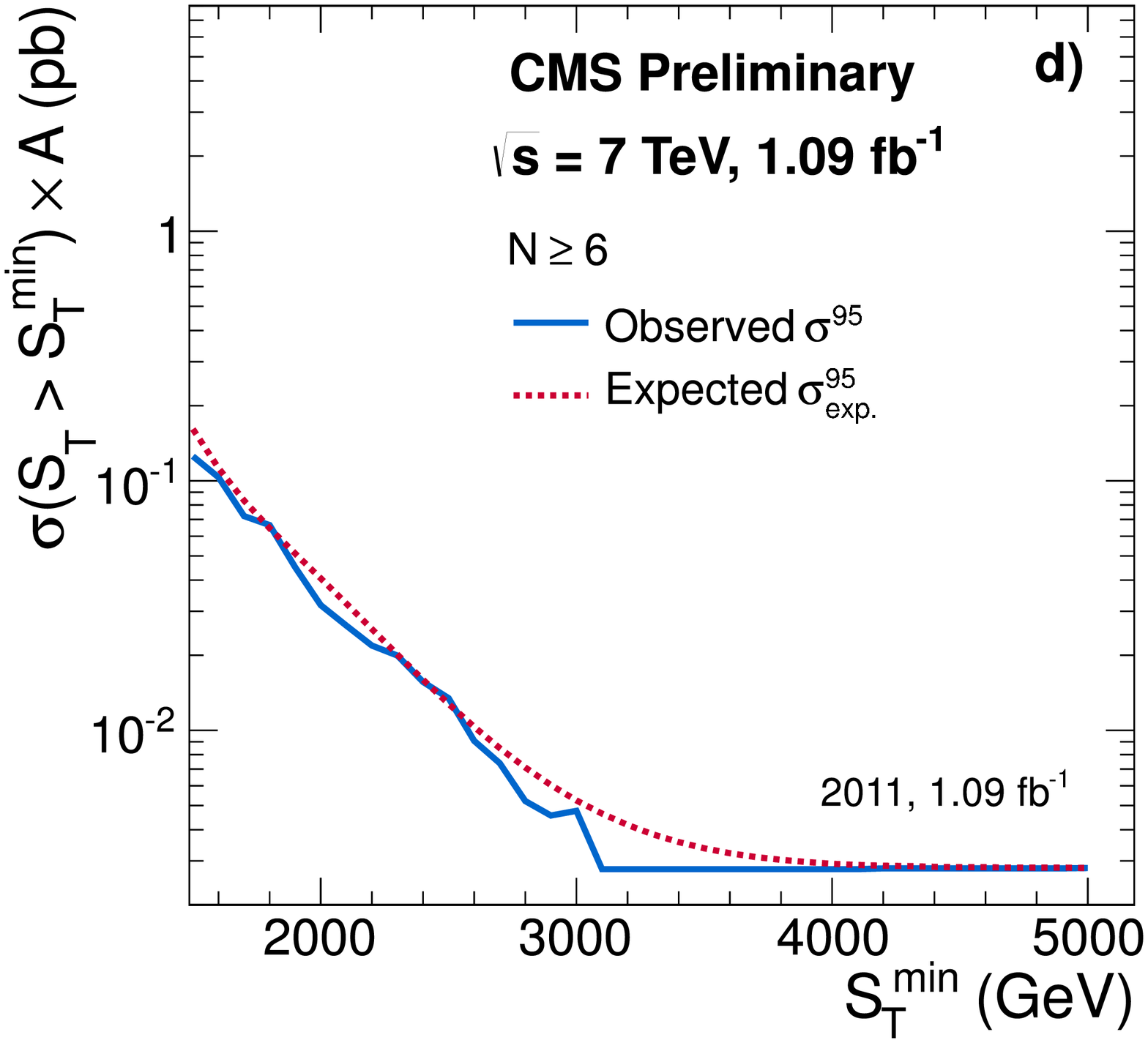}
\caption{a) The 95\% CL limits on black hole model parameters for several benchmark scenarios. The areas below each curve are excluded by this search. b) Cross section limits at 95\% CL on various string ball parameter sets compared with signal production cross section. Colored solid lines show experimental cross section limits, while dotted lines represent corresponding signal cross sections. c, d) Model-independent 95\% CL upper cross section limits for counting experiments with $S_T > S_T^{\rm min}$ as a function of $S_T^{\rm min}$ for c) $N \ge 5$, and d) $N \ge 6$. The blue solid (red dotted) lines correspond to an observed (expected) limit for nominal signal acceptance uncertainty of 5\%.
\label{fig:st_incl}}
\end{figure}

For each inclusive $S_T$ spectrum, counting experiments are performed for $S_T > S_T^{\rm min}$ and model-independent limits on the cross section times the acceptance in high-$S_T$ inclusive final states are produced. These 95\% confidence level (CL) limits can be used to probe for any model or hypothetical particle that results in an energetic high-multiplicity final state. Model-independent limits approach 3~fb at high $S_T$, which is roughly 30 times improvement in sensitivity over the previously published results~\cite{bh_plb}. The model-independent limits for $N \ge 5$ and $N \ge 6$ cases are shown in Fig.~\ref{fig:st_incl}. Finally, 95\% CL limits on the production of black holes and string balls in terms of model parameters that are used in the analysis are computed. All signal samples are produced using the {\sc BlackMax}~\cite{BlackMax} event generator: non-rotating and rotating black holes with $M_D \in [1.5, 3.5]$~TeV, $M_{\rm BH}^{\rm min} \in [3,5.5]$~TeV, and $n = 2, 4,$ and 6; string balls with $M_D \in [1.3,2.1]$~TeV, $M_S \in [1.0,1.7]$~TeV, and $g_S = 0.4$. The Bayesian method with flat signal prior and log-normal prior for integration over background, signal acceptance, and luminosity is used to set 95\% CL limits on the production cross sections. Systematic uncertainties on signal arise from jet energy scale uncertainty (5\%) and parton distribution functions, PDF (2\%). Systematic uncertainty on the background comes mainly from the difference in fit functions shape and varies from 10\% to 290\% depending on the $S_T$ value. A 6\% uncertainty due to luminosity measurements is applied to both signal and background. Translating these limits into the ADD model and {\sc BlackMax} parameters, production of microscopic black holes at the LHC with minimum masses between 4.0 and 5.1~TeV, and string balls with masses 4.1 -- 4.5~TeV is excluded (see Fig.~\ref{fig:st_incl}). These are the most stringent limits to date. The limits on the string balls are the first direct limits on that model set at hadron colliders.  

%%%%%%%%%%%%%%%%%%%%%%%%%%%%%%%%%%
\section{Searches for Extra Dimensions in Diphoton Final State\label{s:diphot}}
As mentioned in Section~\ref{s:intro}, large extra dimensions may result in production of virtual KK graviton modes, that decay to two photons. The total cross section is a superposition of the SM term, direct extra dimensions (ED) term, and interference: $\sigma_{\rm TOT} = \sigma_{\rm SM} + A\eta_{\rm G}\sigma_{\rm int} + B\eta_{\rm G}^{2}\sigma_{\rm ED}$. Here, $\eta_{\rm G} = F/M_{S}^4$, where $M_S$ is the ultraviolet cutoff scale and $F$ is a model parameter for which several conventions exist~\cite{grw,hlz,hewett}.
\begin{eqnarray}
\mathcal{F}&=&1~~~\mbox{(Giudice, Ratazzi, and Wells; GRW \cite{grw})} \nonumber\\
\mathcal{F}& =& \left\{ \begin{array}{cr}
\log(M_{\rm S}^2/\hat{s}) & \mbox{if $n_{\rm ED}=2$} \\
2/(n_{\rm ED}-2) & \mbox{if $n_{\rm ED}>2$}\end{array}\right.\label{eq:HLZ}~~~\mbox{(Han, Lykken, and Zhang; HLZ \cite{hlz})} \nonumber\\
\mathcal{F}&=&\pm\frac{2}{\pi}~~~\mbox{(Hewett \cite{hewett})} \nonumber
\end{eqnarray} 

This search~\cite{add_jhep,exo-10-019} is based on 36~fb$^{-1}$ of $pp$ collision data. Events for the control samples are required to satisfy single-photon triggers requirements with $E_T$ thresholds from 20 to 30~GeV, and events for the signal sample are collected with single-photon trigger and diphoton trigger where each photon must have $E_T$ above 10 -- 22~GeV. Pre-selected events are further required to contain two isolated central ($|\eta| < 1.44$) electromagnetic (EM) clusters with shower shape in the calorimeter consistent with that of a photon and with $E_T >~$30~GeV. The efficiency to reconstruct and identify a pair of such photons is roughly 77\%. Finally, the diphoton mass is required to be $M_{\gamma\gamma} >~$60~GeV. The diphoton mass region between 60 and 200~GeV (200 -- 500~GeV) is considered control (intermediate) region.

Signal samples of ED in the ADD model are simulated using {\sc SHERPA}~\cite{sherpa} event generator for a number of model parameters $M_S$ and $n_{\rm ED}$. Simulation includes both SM and signal diphoton production. Next-to-leading (NLO) order effects are taken into account by applying a flat $K$-factor of 1.3 $\pm$~0.1~\cite{kf_add_diphot}. The RS-1 graviton samples are simulated using {\sc PYTHIA}~\cite{pythia} Monte Carlo (MC) program with mass of the first excited mode $M_1 \in [0.25,2.0]$~TeV and $\tilde{k} = 0.01, 0.05,$ and 0.1. The NLO $K$-factor, applied to signal cross section, depends only weakly on both $\tilde{k}$ and diphoton mass and is taken to be 1.3 $\pm$~0.3~\cite{kf_rs_diphot}. 

The dominant diphoton background is estimated with {\sc SHERPA}, with leading-order cross sections corrected for NLO effects with a $K$-factor of $1.3 \pm 0.3$. Here, 23\% relative systematic uncertainty covers the $K$-factor variation of the diphoton mass. Smaller dijet and photon+jet backgrounds, where jet is misidentified as a photon are estimated from data. First, a rate at which isolated photons are mimicked by non-isolated photons (misidentification rate) is computed from an EM enriched data sample with photon candidates that pass all selection criteria (numerator) and fail isolation or shower shape requirements (denominator). Contamination from real photons resulting from $pp$ collisions is removed in the numerator. The misidentification rate drops from 28\% at $E_T = 30$~GeV to 2\% at $E_T = 120$~GeV. Further, this rate is applied to data samples with one ore more non-isolated photon candidates. A 20\% systematic uncertainty due to the data-driven misidentification rate calculation is applied to the dijet and photon+jet backgrounds.

\begin{figure}[htbp]
\includegraphics[scale=0.28]{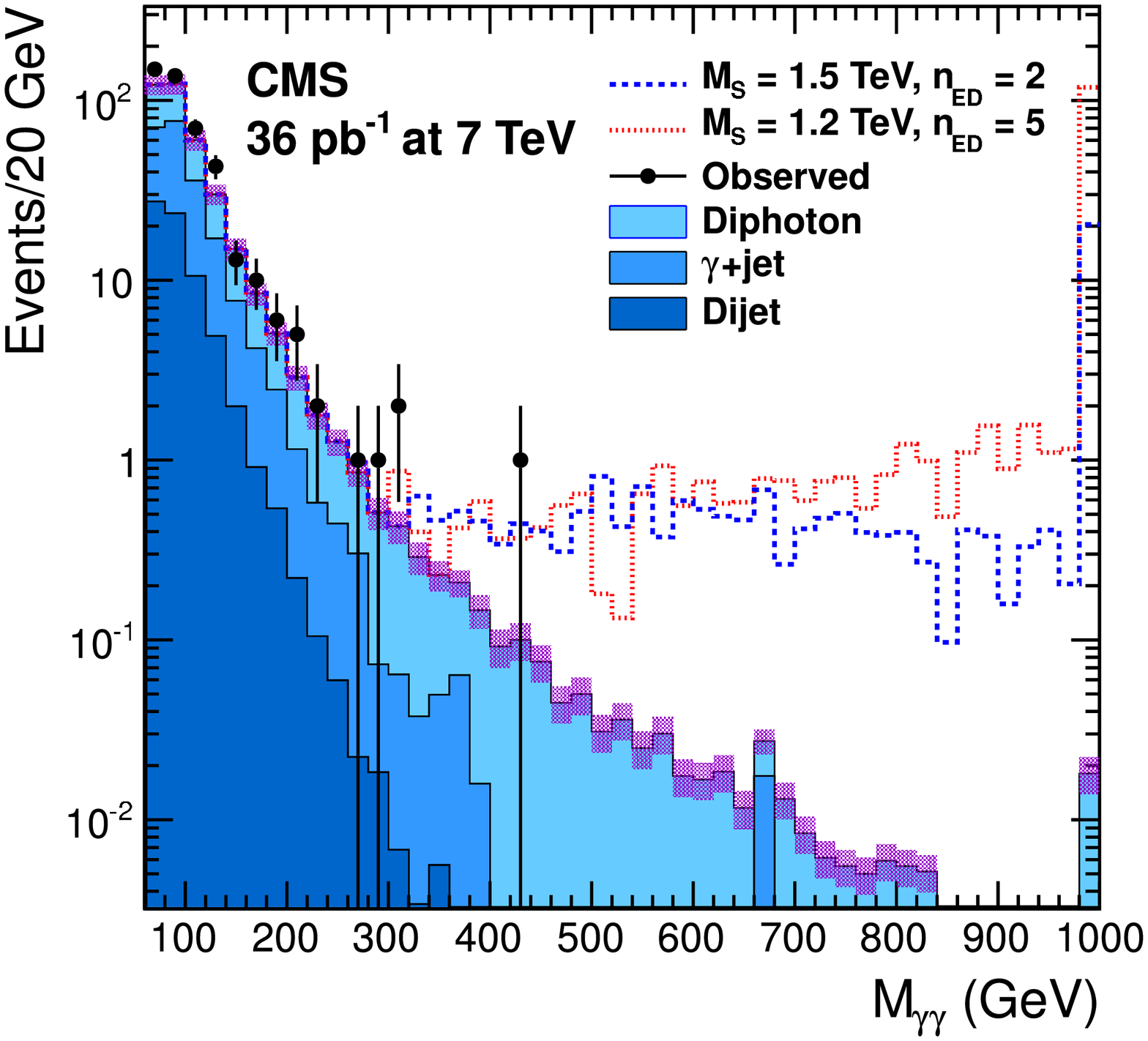}
\includegraphics[scale=0.28]{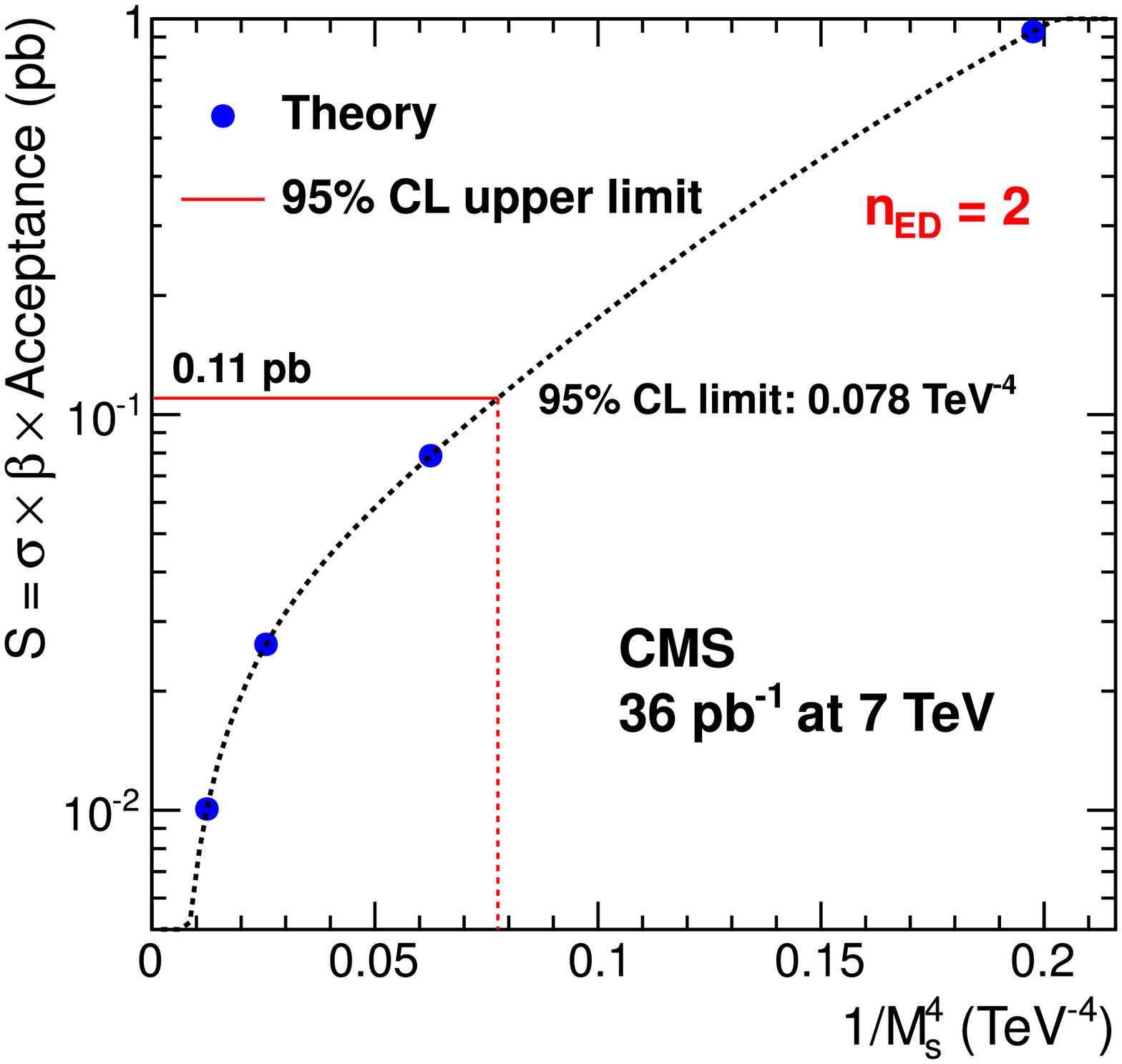}
\includegraphics[scale=0.28]{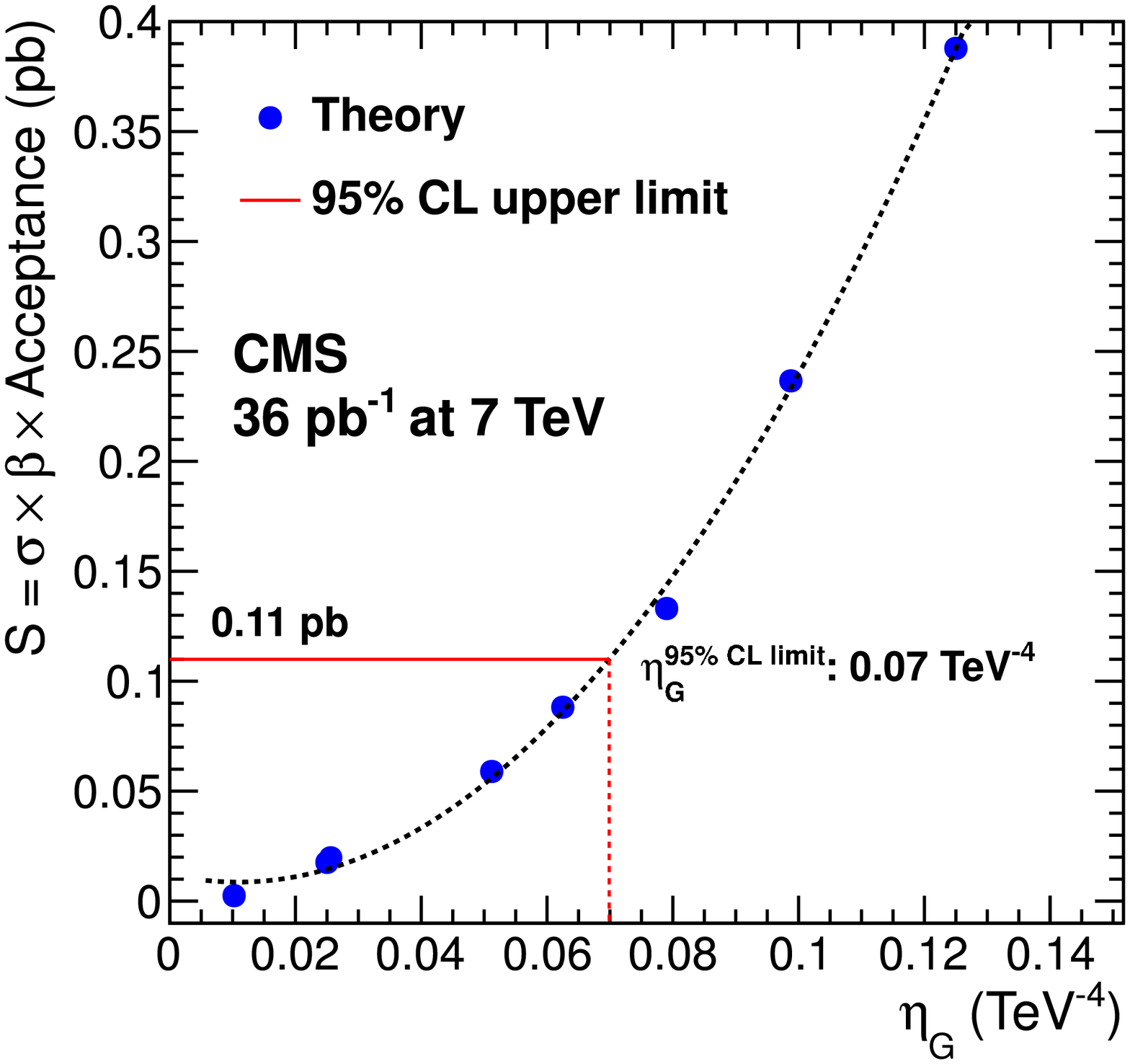}
\caption{(Left) Diphoton mass spectrum in data (black dots), backgrounds (various shades of blue), and two ADD benchmark scenarios (dotted lines). Theoretical cross section times branching ratio times acceptance, $S$, as a function of $1/M_{S}^{4}$ (Middle) and $\eta_G$ (Right). Horizontal red line represents the 95\% confidence level exclusion limits on $S = 0.11$~pb, vertical red lines represent upper limits on model parameters. 
\label{fig:add_diphot}}
\end{figure}

\begin{figure}[htbp]
\includegraphics[scale=0.30]{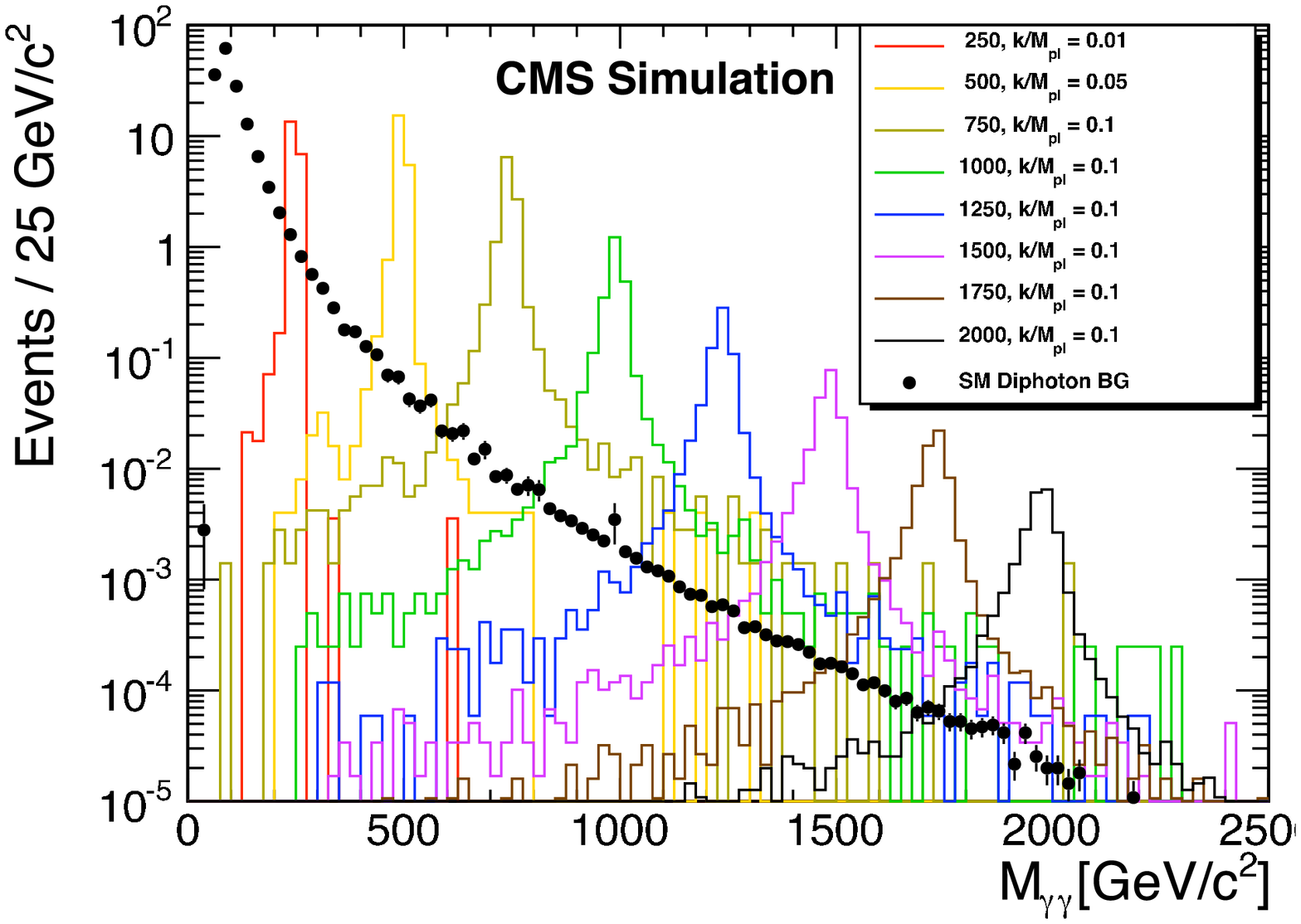}
\includegraphics[scale=0.30]{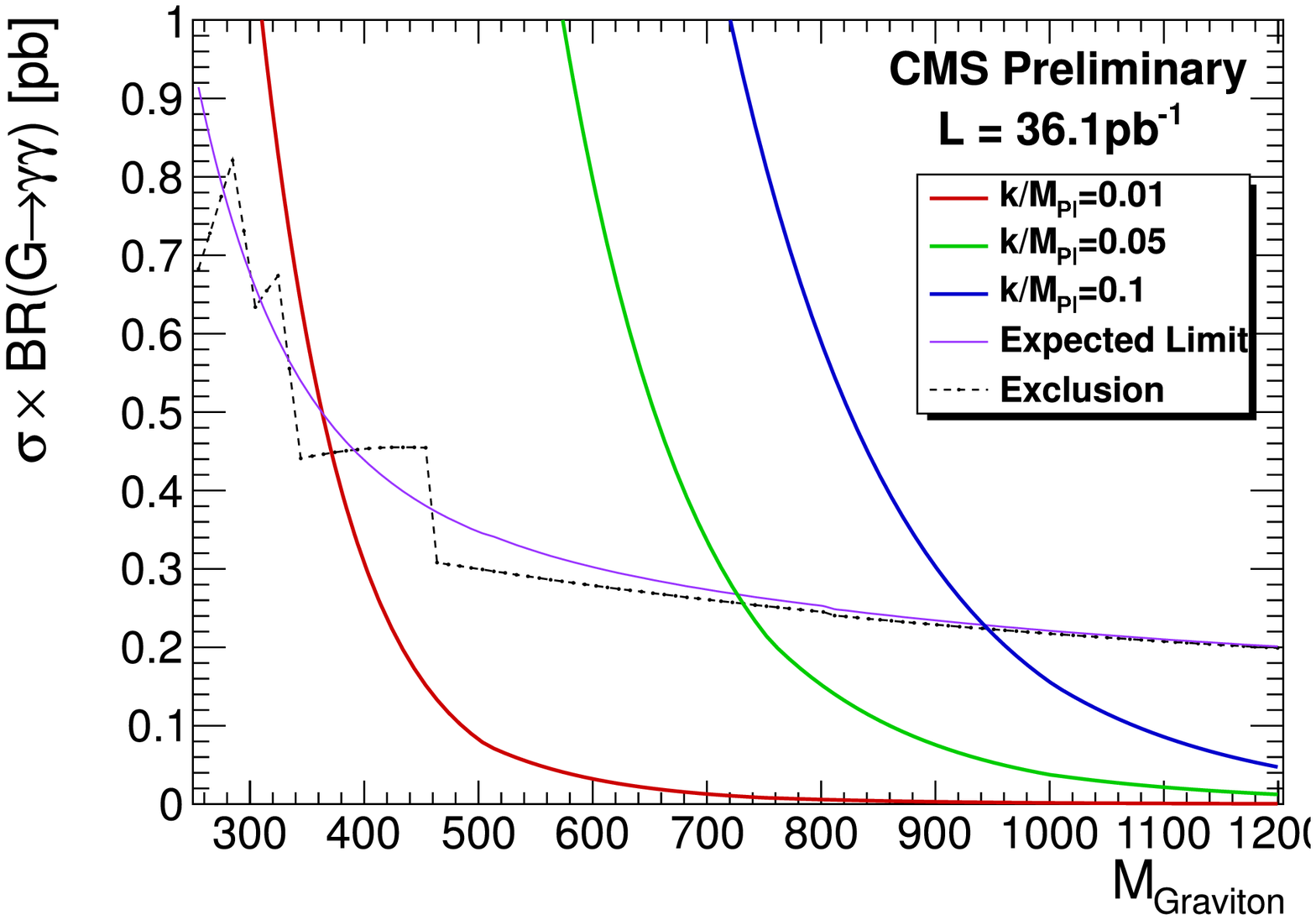}
\includegraphics[scale=0.235]{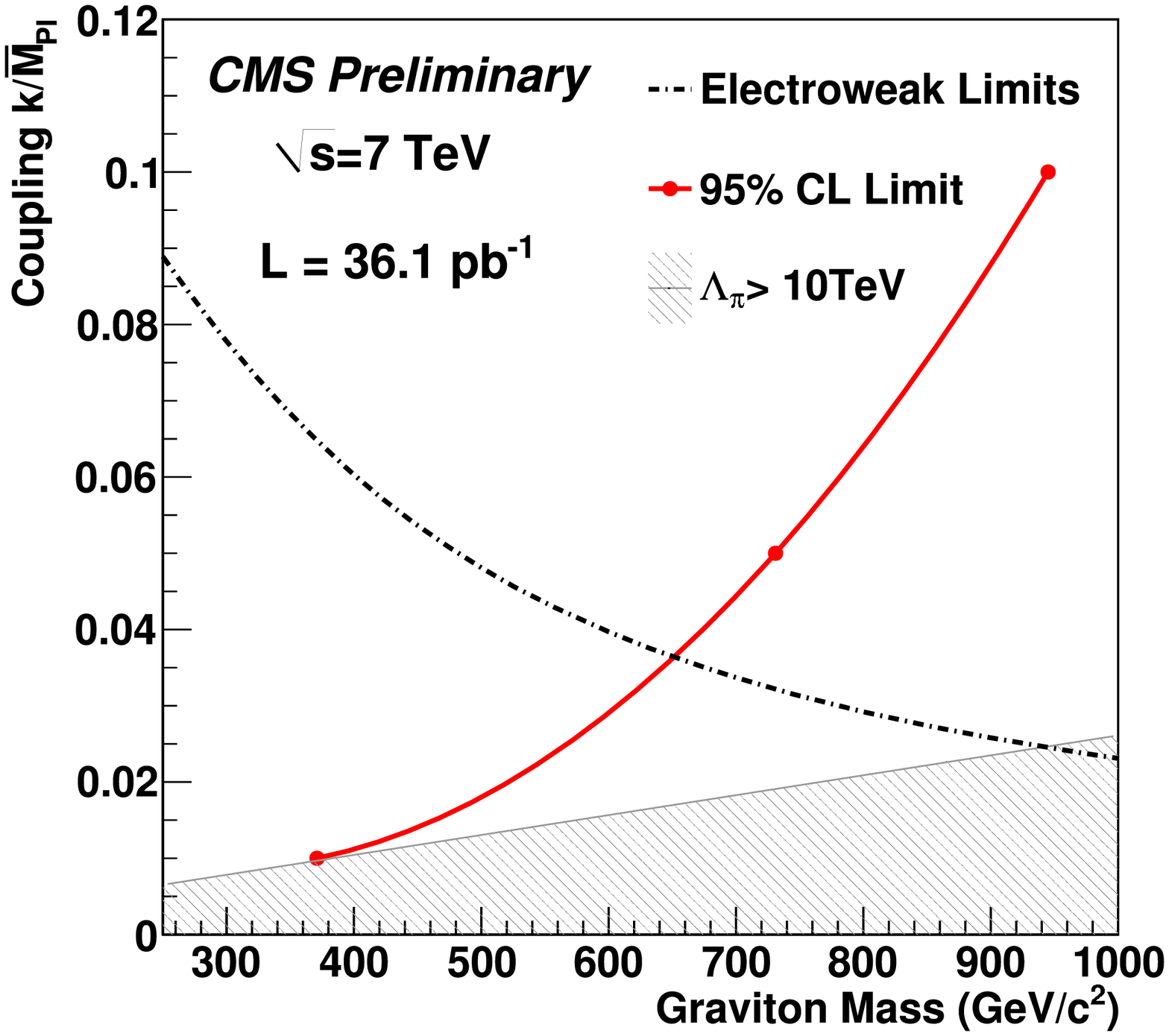}
\caption{(Left) RS graviton diphoton mass distribution compared to SM diphoton background, both normalized to 36~pb$^{-1}$. (Middle) 95\% CL limits on theoretical cross section for three values of coupling $\tilde{k}$. Dashed line represents the exclusion region. (Right) 95\% CL limit (red solid line) on RS graviton mass for different values of the coupling. Area to the left of red line is excluded by this search.
\label{fig:rs_diphot}}
\end{figure}

The diphoton mass spectrum in 36~pb$^{-1}$ of data compared to backgrounds and two ADD benchmark signal samples is shown in Fig.~\ref{fig:add_diphot}. 
Data agree well with background prediction in the control region. In the signal region, zero events is observed which is also consistent with roughly 0.3 expected background events. Due to slight difference in the control samples choice between the ADD and RS analyses, the background prediction numbers differ by less than 1\%. To fit in the page limits of the Proceedings, the diphoton spectrum in data for the RS analysis is not shown (see Ref.~\cite{exo-10-019}). Instead, simulation of RS gravitons normalized to 36~fb$^{-1}$ is shown in Fig.~\ref{fig:rs_diphot}.
   
In the absence of data excess over predicted backgrounds, limits on the model parameters are set. Due to different shape of signal (non-resonant ADD and resonant RS), limit setting procedure differs for two models. For the ADD model, counting experiments are performed in the region with $M_{\gamma\gamma} > 500$~GeV, and 95\% CL limits are set on the cross section times branching ratio times acceptance, $S = (\sigma_{\rm TOT} - \sigma_{\rm SM}) \times \beta \times A$. The Bayesian method is used with flat prior on signal and log-normal prior on nuisance parameters (luminosity, background, and signal efficiency). The systematic uncertainty on the signal efficiency is on the order of 6\%, which accounts for the PDF uncertainty, difference in efficiencies in the MC and data, $etc.$ Roughly 8\% systematic uncertainty comes from the signal diphoton $K$-factor determination. The observed (expected) 95\% CL limit on $S$ is calculated to be 0.11~pb (0.13~pb). Parameterization of $S$ as a function of $1/M_{S}^{4}$ (for $n_{\rm ED} = 2$) and $\eta_G$ are shown in Fig.~\ref{fig:add_diphot}. Red horizontal line on the plots represent the observed limit on $S$, and the $x$-axis value where that line crosses the theoretical parameterization of $S$ marks the 95\% upper limits on the ADD model parameters. 
Consequently, these limits can be translated into lower limits on multidimensional Planck scale, which are summarized in Table~\ref{tab:add_limits_tab}.
Numerically, lower limits on the $M_D$ vary from roughly 1.6 to 2.3~TeV (1.3 to 2.2~TeV) depending on the convention for the full limits (truncated limits, when the signal cross section is set to zero for $\sqrt{\hat s} > M_S$). All limits, except $n_{\rm ED} = 2$ case, are the world's best to the date of Ref.~\cite{add_jhep} publication.

In case of RS gravitons, counting experiments are performed in windows of the diphoton mass. The lower bound of the window is optimized to be $M_{\rm RS} - 5 \times \sigma_{\rm eff}$, where $M_{\rm RS}$ is the mass of the graviton and $\sigma_{\rm eff}$ is the half-width of the narrowest mass interval that contains 68\% of the signal. The upper bound on the window is set to infinity. The background diphoton mass distribution is parameterized by a fit to MC in the region of $220 < M_{\gamma\gamma} < 1000$~GeV, normalized to the data in $100 < M_{\gamma\gamma} < 220$~GeV, and extrapolated into the signal region. The rest of the limit setting procedure is identical to that of the ADD analysis. The 95\% CL limits on the theoretical cross section times branching ratio are shown in Fig.~\ref{fig:rs_diphot} as a function of the first excited mode for three different coupling values. Also shown is the exclusion plot as a function of $M_1$. This search excludes RS gravitons with masses below 371, 731, and 945~GeV for couplings $k/M_{Pl} = 0.01$, 0.05, and 0.1, respectively.

\begin{table}[htbp]
\centering
\caption{95\% CL lower limits on multidimensional Planck scale in the ADD model framework for different conventions (Giudice, Ratazzi, and Wells (GRW); Hewett; Han, Lykken, and Zhang (HLZ)). Truncated limits are set assuming signal cross section is set to zero for $\sqrt{\hat s} > M_{S}$.\label{tab:add_limits_tab}}
\begin{tabular}{|c|c|cc|cccccc|}
\hline
& $\bm {M_D}$,~\textbf{TeV (GRW)} & \multicolumn{2}{c|}{$\bm {M_D}$,~\textbf{TeV (Hewett)}} & \multicolumn{6}{c|}{$\bm {M_D}$,~\textbf{TeV (HLZ)}} \\
& & Pos. & Neg. & $n_{\rm ED}=2$ & $n_{\rm ED}=3$ & $n_{\rm ED}=4$ & $n_{\rm ED}=5$ & $n_{\rm ED}=6$ & $n_{\rm ED}=7$ \\\hline
Full & 1.94 & 1.74 & 1.71 & 1.89 & 2.31 & 1.94 & 1.76 & 1.63 & 1.55 \\
Trunc. & 1.84 & 1.60 & 1.50 & 1.80 & 2.23 & 1.84 & 1.63 & 1.46 & 1.31\\
\hline
\end{tabular}
\end{table}
%%%%%%%%%%%%%%%%%%%%%%%%%%%%%%%%%%
\section{Searches for Extra Dimensions in Dimuon Final State\label{s:dimu}}
The search for ADD extra dimensions in the dimuon final state~\cite{exo-10-020} is based on 39.7~pb$^{-1}$ of data collected by the CMS in 2010. Pre-selection is based on single muon triggers with lowest $p_T$ thresholds varying from 9 to 15~GeV, that are 99\% efficient in the control and search regions. Further, at least two isolated muon candidates with $p_T > 30$~GeV and with hits in both tracker and muon system are required to be present in the event. Muon candidates should not be exactly back-to-back to suppress background from cosmic muons.

\begin{figure}[htbp]
\includegraphics[scale=0.28]{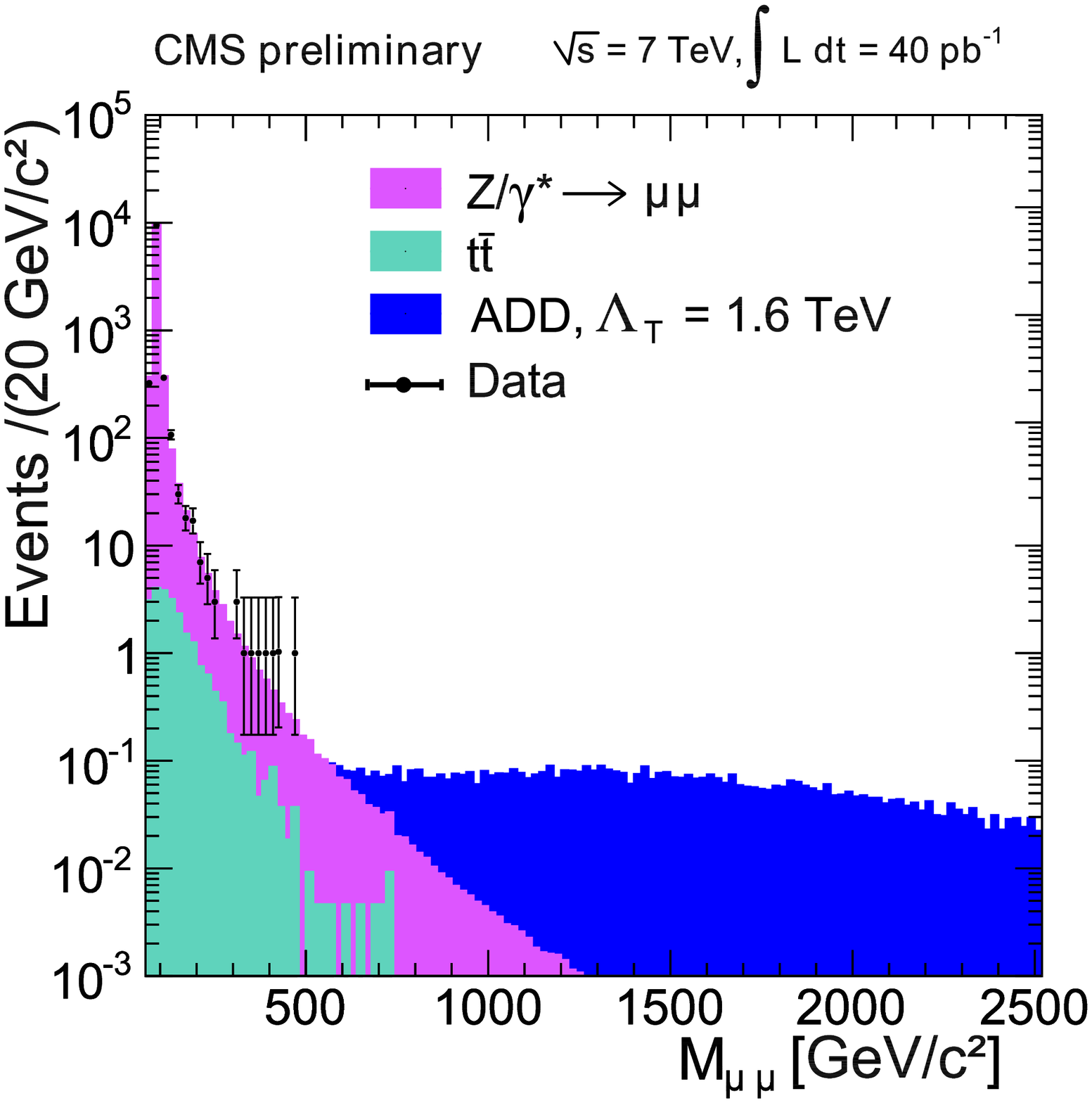}
\includegraphics[scale=0.28]{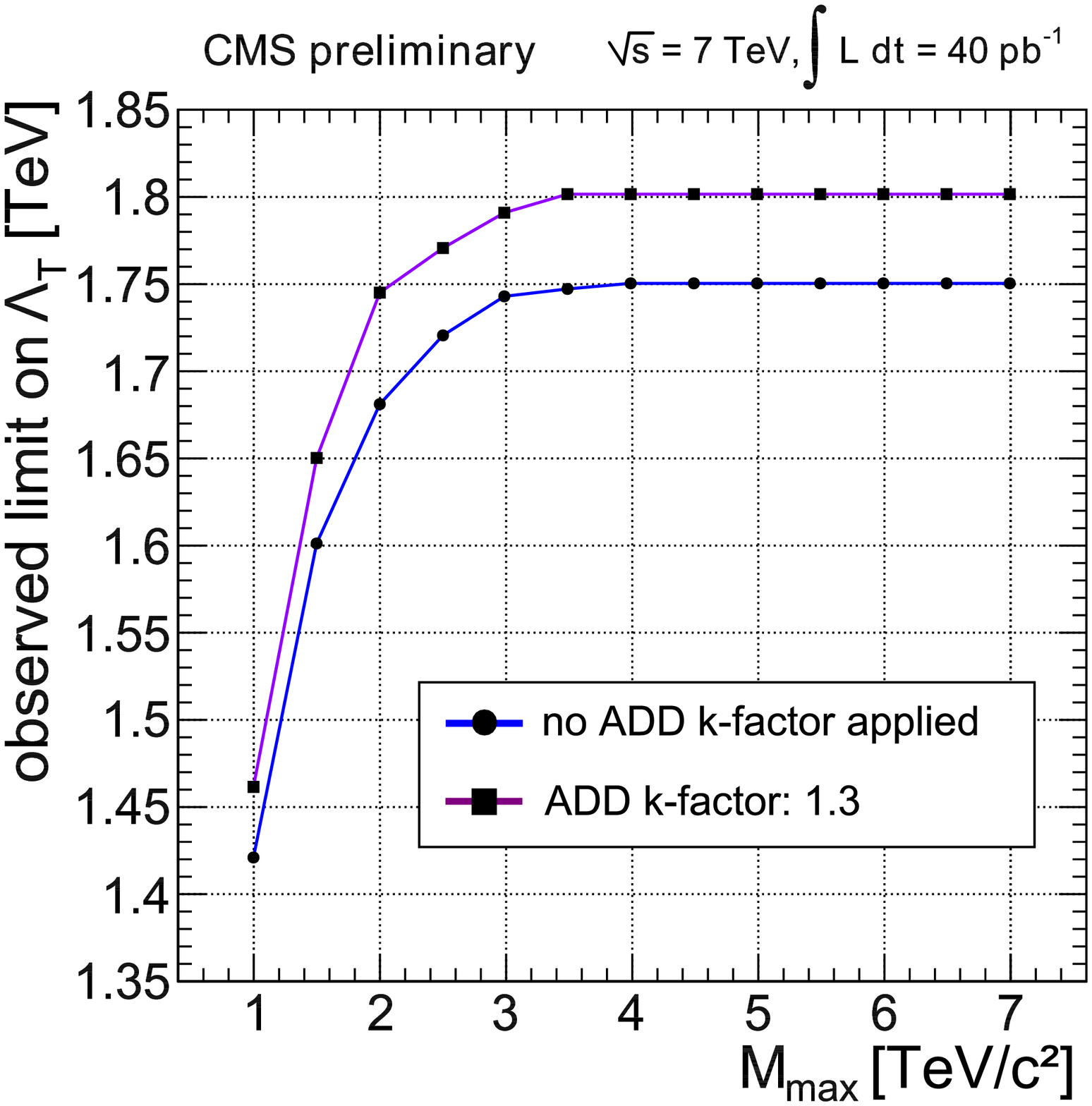}
\includegraphics[scale=0.28]{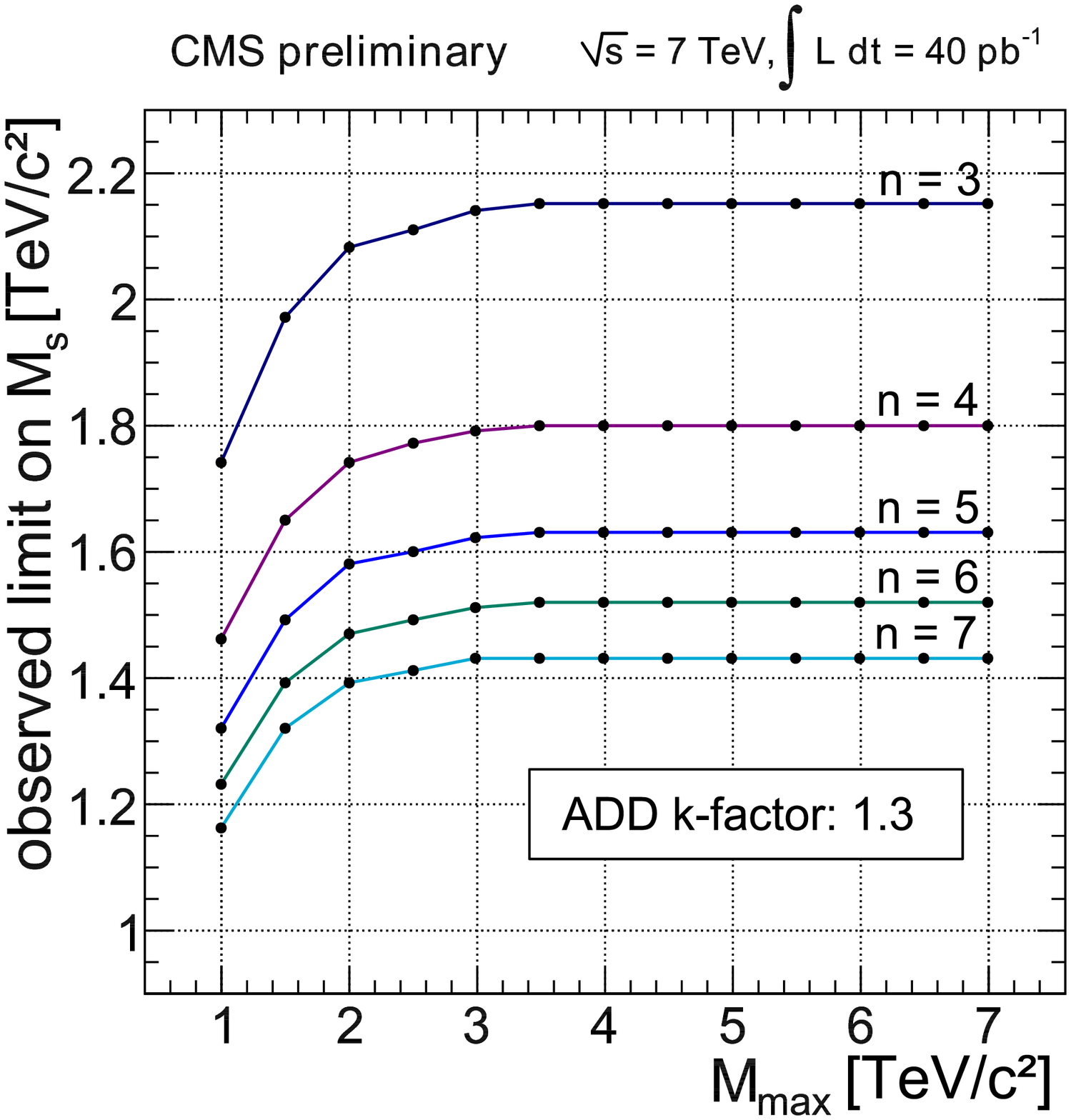}
\caption{(Left) Dimuon mass spectrum in data (dots) compred to background prediction (pink and green histograms) and benchmark ADD model (blue histogram). (Middle) 95\% CL observed upper limits on $\Lambda_T$ with and without NLO corrections applied to signal cross section. (Right) Observed 95\% CL upper limits on $M_S$ for different numbers of extra dimensions.
\label{fig:add_dimuon}}
\end{figure}

The {\sc PYTHIA} leading order (LO) event generator is used to simulate ADD signal with $\Lambda_T \in [800, 1800]$~GeV~\cite{grw}. Here, $\Lambda_T$ is the model parameter that describes the model phenomenology after summation over KK towers: $\Lambda_T^4 = \frac{8 \pi \, \Gamma \left( n / 2 \right)}{2 \pi^{n / 2} c_1 } \cdot \frac{M_D^{n+2}}{\Lambda^{n-2}}$, where $\Lambda$ is the ultraviolet cutoff that eliminates the summation divergence. Both SM and virtual graviton effects are described by the generator. The dominant Drell-Yan background is also estimated at LO using {\sc PYTHIA} simulation. The NLO and next-to-next-to-leading order (NNLO) effects have been studied in detail using various higher orders generators, and are found to be 1.4 near the $Z$ boson peak position and 1.45 at $M_{\mu\mu} > 600$~GeV. The $t \bar t$ background is estimated using simulation and cross-checked with the latest CMS $t \bar t$ cross section measurement. QCD background is found to be negligible after the offline cuts. The resulting dimuon mass spectra in data and backgrounds are shown in Fig.~\ref{fig:add_dimuon}, compared to the benchmark ADD scenario. The agreement between data and the backgrounds is satisfactory in the full range of masses. No excess of data over the backgrounds is observed. Higher order corrections result in 15\% systematic uncertainty (dominant) on the background, followed by 5\% Drell-Yan PDF uncertainties and 4\% muon momentum resolution. The largest systematic uncertainty on signal is trigger and reconstruction uncertainty of 4\%. Luminosity uncertainty is 6\% for both signal and background. 

Counting experiments are performed to set limits on the signal cross sections, $\sigma_S$, in the dimuon mass spectrum above the optimal cut value of 600~GeV. This threshold has been optimized with respect to the expected limit on $\Lambda_T$ in the full range of model validity ({\it i.e.} up to 7~TeV). The Bayesian method is used with flat signal prior. The 95\% observed upper limits on $\sigma_S$ are calculated to be between 0.088 and 0.098~pb. These limits can be translated into limits on the ADD model parameters in different conventions. Figure~\ref{fig:add_dimuon} shows 95\% CL observed upper limits on GRW convention parameter $\Lambda_T$ as a function of maximal mass of a muon pair for the LO approximation and with NLO $K$-factor of 1.3~\cite{kf_add_dimu1,kf_add_dimu2} applied to the signal cross section. Also shown are the 95\% CL limits on $M_S$ (HLZ convention) for different numbers of extra dimensions. Numerically, these limits are summarized in Table~\ref{tab:add_limits_dimu}. The observed limits on ADD models for number of extra dimensions above two are the best limits set at hadron colliders in the dimuon final state.

\begin{table}[htbp]
\centering
\caption{95\% CL lower limits on $\Lambda_T$ and $M_S$ in the ADD model framework for different conventions. Truncated limits are set assuming signal cross section is set to zero for $\sqrt{\hat s} > M_{S}$.\label{tab:add_limits_dimu}}
\begin{tabular}{|c|c|cccccc|}
\hline
& $\bm {\Lambda_T}$,~\textbf{TeV (GRW)} & \multicolumn{6}{c|}{$\bm {M_S}$,~\textbf{TeV (HLZ)}} \\
& & $n_{\rm ED}=2$ & $n_{\rm ED}=3$ & $n_{\rm ED}=4$ & $n_{\rm ED}=5$ & $n_{\rm ED}=6$ & $n_{\rm ED}=7$ \\\hline
Full & 1.80 & 1.75 & 2.15 & 1.80 & 1.63 & 1.52 & 1.43 \\
Trunc. & 1.68 & 1.67 & 2.09 & 1.68 & 1.49 & 1.34 & 1.24\\
\hline
\end{tabular}
\end{table}

%%%%%%%%%%%%%%%%%%%%%%%%%%%%%%%%%%
\section{Conclusions}

In conclusion, searches for large extra dimensions have been performed in multi-particle~\cite{exo-11-071}, diphoton~\cite{add_jhep,exo-10-019}, and dimuon~\cite{exo-10-020} final states using up to 1.1~fb$^{-1}$ of 7~TeV center-of-mass $pp$ collision data collected with the CMS detector. Data are found to be in good agreement with the background predictions -- no significant excess of data is observed in any of the four described analyses. Stringent model-independent limits on the production of energetic multi-object final states, as well as limits on ADD and RS model parameters are set. Most of these limits are the best limits set at hadron colliders to date. 

\begin{acknowledgments}
The author would like to thank his CMS colleagues for the material and results presented in this Letter and the LHC for delivering the high quality data that made these searches possible.  
\end{acknowledgments}

\bigskip % extra skip inserted

\end{document}